\documentclass[usletter, 10pt, conference]{cssconf}

\IEEEoverridecommandlockouts 

\usepackage{graphics}
\usepackage{amsmath} 
\usepackage{amssymb}  
\usepackage{mathptmx}
\usepackage{bm}
\usepackage{graphicx}
\usepackage{psfrag}

\newtheorem{thm}{Theorem}%
\newtheorem{rmk}{Remark}

\newcommand{\bket}[1]{\left<#1\right>}
\newcommand{\ket}[1]{\left|#1\right>}
\newcommand{\bra}[1]{\left<#1\right|}
\newcommand{\tr}[1]{\text{Tr}\left(#1\right)}
\newcommand{\dotex}{\frac{d}{dt}}

\begin{document}
\bibliographystyle{plain}

\title{ Parameter estimation of a  $3$-level quantum system with a single population measurement   \thanks{This work was   supported in part by the "Agence Nationale de la Recherche" (ANR),
    Projet Blanc  CQUID number 06-3-13957.}}

\author{
  Zaki Leghtas\thanks{ Mines ParisTech,
 60 Bd Saint-Michel, 75272 Paris cedex 06, FRANCE.
  Email: zaki.leghtas@ensmp.fr}
  \and
  Mazyar Mirrahimi\thanks{INRIA Rocquencourt, Domaine de Voluceau,
 Rocquencourt B.P. 105, 78153 Le Chesnay Cedex, FRANCE.
  Email: mazyar.mirrahimi@inria.fr}
 \and
  Pierre Rouchon\thanks{Mines ParisTech, Centre Automatique et Syst\`{e}mes, Math\'{e}matiques et Syst\`{e}mes,
 60 Bd Saint-Michel, 75272 Paris cedex 06, FRANCE.
  Email: pierre.rouchon@mines-paristech.fr}
  }

\maketitle \thispagestyle{empty} \pagestyle{empty}

\begin{abstract}
An observer-based  Hamiltonian  identification algorithm
            for quantum systems has been proposed in \cite{bonnabel-et-al:auto09}. The
            later paper provided a method to estimate the dipole moment matrix of a quantum system requiring the measurement of the
            populations on all states,  which could be   experimentally
             difficult to achieve. We propose here
             an extension to a $3$-level   quantum system,  having
            access to the  population of the ground state  only. By a
            more adapted choice of  the control field, we will
            show that a  continuous measurement of this  observable,
            alone, is enough to identify the field coupling parameters (dipole moment).

\end{abstract}


\section{Introduction}
For applications ranging from quantum computers to the synthesis of new molecules, an accurate estimation of the parameters involved in the dynamics of the quantum system is fundamental.
Various methods have been engineered over the years like the maximum-likelihood methods \cite{kosut-et-al:ifac03,kosut-et-al:arxiv04a,kosut-et-al:arxiv04b}, the maximum-entropy methods \cite{buzek:LN04} and minimum Kullback entropy methods \cite{olivares-paris:PhRevA07}. The optimal identification techniques via
least-square criteria's \cite{geremia-rabitz:JCPh03,geremia-rabitz:PRL02,bris-et-al:cocv07} and the map inversion techniques
\cite{shenvi-et-al:JPC02} are some other techniques explored in this area.
In \cite{kosut-rabitz:ifac02}, a state-observer is presented for the state identification combined with a gradient method on the dipole moment. This result was then improved in \cite{bonnabel-et-al:auto09} succeeding in simultaneously estimating the state of the system and it's dipole moment using observers. In~\cite{bris-et-al:cocv07} a rigorous proof of the well-posedness of the problem is proposed. All these results required the knowledge of the populations on all energy levels. Experimentally, that is extremely difficult to achieve. Since in quantum mechanics, measuring an observable influences the system, the less information we need, the less we disturb the system, and the more likely our estimation is accurate.
Our goal was to improve the result given in \cite{bonnabel-et-al:auto09} in order to   estimate the dipole moments of a quantum system measuring continuously the population on the first state only. We focus here on $3$-level  systems with a single population measure.

In section~\ref{sec:model}, we explain the $3$-level system and set the  estimation problem attached to~\eqref{eq:dyn}. In section~\ref{sec:sim} we present a $2$-step estimation procedure based on two nonlinear asymptotic observers~\eqref{eq:obs12} and~\eqref{eq:obs23}. Sections~\ref{sec:12} and~\ref{sec:23} are devoted to local convergence proofs.

\section{The $3$-level system} \label{sec:model}

\subsection{Model and  problem setting}
Denote by $\ket{k}$, $k=1,2,3$ the $3$ states of energies $E_k$ such that $|E_2-E_1| \neq |E_3-E_2|$.   Throughout  the paper we use the following notations for $k,l=1,2,3$:
$\sigma^{lk}=\ket{l}\bra{k}-\ket{k}\bra{l}$,  $\sigma_x^{lk}=\ket{l}\bra{k}+\ket{k}\bra{l}$,  $\sigma_z^{lk}=\ket{l}\bra{l}-\ket{k}\bra{k}$ and $P_k=\ket{k}\bra{k}$ (projector on $\ket{k}$).
Assume that the dynamics is described by the following Schr\"{o}dinger equation:
$$
\dotex\ket{\Psi}=\frac{-\imath}{\hbar}(H_0+A(t)H_1)\ket{\Psi},\qquad y=\bket{\Psi|P_1|\Psi},
$$
where $\ket{\Psi}$ is the wave-function,
$A(t)\in\mathbb{R}$ the electromagnetic  field, $H_0=\sum_{k=1}^{3} E_k P_k$  the free Hamiltonian,
$H_1=\mu_{12} \sigma_x^{12} +\mu_{23} \sigma_x^{23}$  the   Hamiltonian matrix describing the coupling with the electromagnetic field (dipole moment) and $y$ the measurement output.
 Assuming the energies $E_k$ known,  the goal consists in estimating the real coupling parameters  $\mu_{12}$ and $\mu_{23}$ from the output $y$.
 We assume the  electromagnetic field  resonant with transitions $1-2$ and $2-3$:
 $$
 A(t)= u_{12} \bar A_{12}\sin\left(\frac{E_2-E_1}{\hbar}t\right)+u_{23} \bar A_{23}\sin\left(\frac{E_3-E_2}{\hbar}t\right)
 $$
 with  small  amplitude magnitudes  $ \bar A_{12}$ and $ \bar A_{23}$ and normalized slow modulations $|u_{12}|, |u_{23}|\in[0,1]$. We have
 $|\bar A_{12}\mu_{12}|, |\bar A_{23}\mu_{23}| \ll |E_2-E_1|, |E_3-E_2|$. In the interaction frame $\ket{\Phi} =e^{\frac{\imath H_0}{\hbar}t} \ket{\Psi}$ and after neglecting highly oscillating terms (rotating wave approximation) we get  the following model
$$
\dotex \ket{\Phi} =  \left( u_{12}\Omega_{12} \sigma^{12} + u_{23}\Omega_{23}  \sigma^{23}\right) \ket{\Phi}
$$
where  $\Omega_{12}=\frac{\bar{A}_{12}\mu_{12}}{2\hbar}$ and  $\Omega_{23}=\frac{\bar{A}_{23}\mu_{23}}{2\hbar}$ are  Rabi amplitudes when $(u_{12},u_{23})=(1,0)$ and $(u_{12},u_{23})=(0,1)$.

In the sequel we will use the density operator $\rho=\ket{\Phi}\bra{\Phi}$ instead of the wave function  $\ket{\Phi}$.  The estimation of  the real parameters $\mu_{12}$ and $\mu_{23}$ is then equivalent to estimation of the two other real parameters   $\Omega_{12}$ and $\Omega_{23}$  appearing in the dynamics of the projector $\rho$
\begin{equation}\label{eq:dyn}
\dotex \rho = u_{12} \Omega_{12}  \left[\sigma^{12}, \rho\right]
+u_{23} \Omega_{23}  \left[ \sigma^{23}, \rho\right]
\end{equation}
via the output $y=\tr{P_1\rho}$ and using  $u_{12}$ and $u_{23}$ as  excitation real inputs.     Remember that $\sigma^{12}$ and $\sigma^{23}$ are anti-symmetric and real matrices: if  the entries of  $\rho$ are  initially real,  they remain real; if $\rho$ is initially  a projector and thus describes a pure quantum  state, its  remains a  projector.  Since we are in the $3$-level case, $\rho$ can be seen as a point on the two dimensional manifold $\mathbb{RP}^2$, the projective space.

\subsection{Identifiability}
It is  proved in \cite{bris-et-al:cocv07} that it is possible to identify  $\Omega_{12}$ and $\Omega_{23}$ by measuring all the populations, i.e., via the measurement outputs $(\tr{P_1\rho},\tr{P_2\rho})$ ($\tr{P_3\rho}=1-\tr{P_1\rho}-\tr{P_2\rho}$). With just $y=\tr{P_1\rho}$, we provide here below arguments showing identifiability via adapted choices  for inputs $u_{12}$ and $u_{23}$.   With $u_{12}=1$ and $u_{23}=0$ we recover essentially a $2$-level system with states $\ket{1}$ and $\ket{2}$  and we can identify $\Omega_{12}$ from $y$ following~\cite{bonnabel-et-al:auto09}. This corresponds to the first step that is treated in theorem~\ref{thm:12}.

Assume now that $\Omega_{12}$ is known. Set
$u_{12}=1$ and   $u_{23}=\eta \cos\theta $ with $\dotex \theta = \Omega_{12}$ and $\eta$ a small positive parameter. With  $\xi =e^{-\theta\sigma^{12}}\rho e^{\theta\sigma^{12}} $, the output map becomes
{\small
$$
y(t)=\frac{\tr{(P_1+P_2)\xi}+\cos(2\theta)\tr{\sigma_z^{12}\xi}
+\sin(2\theta)\tr{\sigma_x^{12}\xi}}{2}
$$}
 and $\xi$ obeys  to
$$
\dotex{\xi}=\eta\cos^2\theta ~\Omega_{23}[\sigma^{23},\xi]+
\eta\cos\theta\sin\theta~\Omega_{23}[\sigma^{23},\xi]
$$
Since $\eta\ll 1$ we can average  its  dynamics:
\begin{equation}\label{eq:av23}
\dotex{\xi}=\frac{\eta\Omega_{23}}{2}[\sigma^{23},\xi]
.
\end{equation}
The average values of $ y(t)(1+2\cos (2\theta))$ and $y(t)(1-2\cos(2\theta ))$ are  $\tr{P_1 \xi}$ and  $\tr{P_2 \xi}$, respectively. Thus in average all the  populations are measured  and according to \cite{bris-et-al:cocv07},  $\Omega_{23}$ is identifiable. This second step is treated in theorem~\ref{thm:23}.

\section{Estimation algorithms  and simulations} \label{sec:sim}
As explained here above, we proceed in two step. In a first step we set in~\eqref{eq:dyn}, $u_{12}=1$ and $u_{23}=0$  and estimate from  the output $y(t)$ the parameter $\Omega_{12}$ via the following nonlinear dynamical system (an invariant nonlinear observer inspired by~\cite{bonnabel-et-al:auto09,bonnabel-et-al:IEEE08,bonnabel-et-al:ifac08a}):
{\small
\begin{align}
&\dotex{\hat\rho}=\hat\Omega_{12}[\sigma^{12},\hat\rho]+ ... \notag
\\&...\epsilon\Gamma_{12}(y(t)-\tr{P_1\hat\rho})(\sigma_z^{12} \hat\rho+\hat\rho \sigma_z^{12}-2\tr{\sigma_z^{12}\hat\rho}\hat\rho)\notag
\\
&\dotex{\hat{\Omega}}_{12}=\epsilon^2\gamma_{12}
     \tr{\sigma_z^{12}[\sigma^{12},\hat\rho]} (y(t)-\tr{P_1\hat\rho}) \label{eq:obs12}
\end{align}}
with $\Gamma_{12},\gamma_{12}$ positive parameters of order~1 and $\epsilon$ a small positive parameter. Local convergence is proved in theorem~\ref{thm:12}. The dynamics~\eqref{eq:obs12} respect two important features: if  the entries $\hat\rho$  are initially real,  they remain real for  $t>0$; if $\hat\rho$ is  initially a  projector and thus describes a pure quantum  state, it remains  a projector for  $t>0$.

Assuming $\Omega_{12}$ obtained via this first step, we take, as explained in previous section,  $u_{12}=1$ with   $u_{23}=\eta \cos\theta $ ($\dotex \theta = \Omega_{12}$ and $\eta$ a small positive parameter) to estimate $\Omega_{23}$ via a second nonlinear dynamical system
{\small \begin{align}
&\dotex{\hat\rho}=\Omega_{12}[\sigma^{12},\hat\rho]+\eta\cos\theta~\hat\Omega_{23}[\sigma^{23},\hat\rho]+...
\notag
\\
&...\epsilon\eta\Gamma_{23}(y-\hat y)(1-2\cos(2\theta))\left(\Sigma_z^{23}\hat\rho+\hat\rho \Sigma_z^{23}-2\tr{\Sigma_z^{23}\hat\rho}\hat\rho\right) \notag
\\
&\dotex{\hat{\Omega}}_{23}=\epsilon^2\eta\gamma_{23}(y-\hat y)(1-2\cos(2\theta))\tr{\Sigma_z^{23}[\Sigma^{23},\hat\rho]} \label{eq:obs23}
\end{align}}
where
$\Sigma^{23}=U(t)\sigma^{23}U^\dag(t)$, $\Sigma_z^{23}=U(t)\sigma_z	^{23}U^\dag(t)
$  with $U(t)=\exp(\theta(t)\sigma^{12})$ and where $\Gamma_{23}$ and $\gamma_{23}$ are positive parameters of order~1 and $\epsilon$ is a small positive parameter. Local convergence  is addressed in theorem~\ref{thm:23}.
\begin{figure}[htb]
  \centerline{\includegraphics[width=0.5\textwidth]{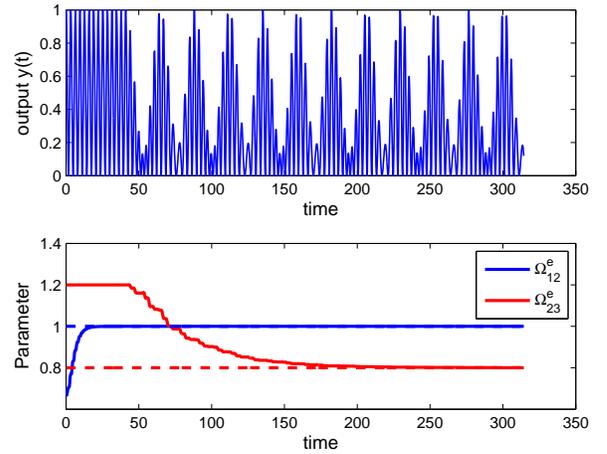}}
  \caption{Estimation of $\Omega_{12}$ in a first step ($t\in[0,50]$) via~\eqref{eq:obs12}; estimation of $\Omega_{23}$ in a second step ($t>50$) via~\eqref{eq:obs23} (no modeling  and measure errors, $\Omega^e_{12}$ and $\Omega^e_{23}$ stand for $\hat\Omega_{12}$ and $\hat\Omega_{23}$). }\label{fig:perfect}
\end{figure}

\begin{figure}[htb]
  \centerline{\includegraphics[width=0.5\textwidth]{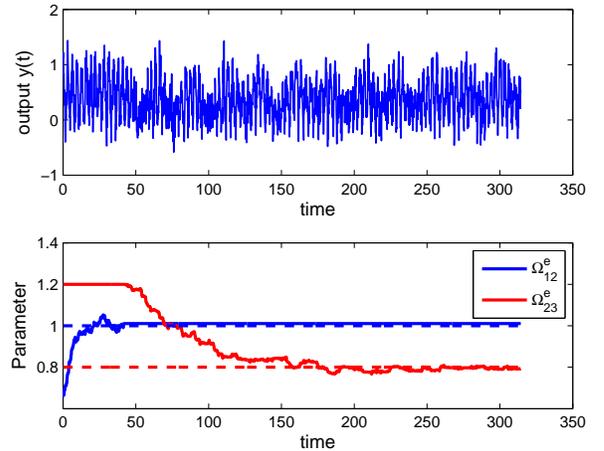}}
  \caption{Similar simulations to those of  figure~\ref{fig:perfect} but with $20\%$ of Gaussian  additive noise on the output y(t) and $10\%$ of  Gaussian additive noise on the inputs $u_{12}$ and $u_{23}$.}\label{fig:noise}
\end{figure}

Let us look at some simulations (figures~\ref{fig:perfect} and~\ref{fig:noise}) with the following numerical values:  $\rho(0)=\hat\rho(0)=P_1$, $\Omega_{12}=1.0$, $\Omega_{23}=0.8$, $\hat\Omega_{12}(0)=\frac{\Omega_{12}}{1.5}$, $\hat\Omega_{23}(0)=1.5 \Omega_{23}$, $\epsilon=\eta=\frac{1}{3}$, $\Gamma_{12}=\Gamma_{23}=4$, $\gamma_{12}=\gamma_{23}=1$.   Convergence of $\hat\Omega_{12}$ is effective after $t>20$ that  corresponds to  $\frac{1}{\epsilon}$ Rabi periods associated to  transition $1-2$ since $\frac{2\pi}{\epsilon\Omega_{12}}\approx 18$. Convergence of $\hat\Omega_{23}$ is achieved for $t\in[50,200]$. The interval length  corresponds to $\frac{1}{\epsilon\eta}$ Rabi periods associated to  transition $2-3$ for the average dynamics~\eqref{eq:av23} since  $\frac{4\pi}{\epsilon\eta\Omega_{23}}\approx 140 $. These convergence times are in good agreement with the convergence times  that can be obtained from the linearized  system~\eqref{eq:linav12}  appearing during the proof of theorem~\ref{thm:12}. When additive noises are introduced on the inputs and output, the performance are not dramatically changed and the convergence times are almost the same.

\section{Estimation of $\Omega_{12}$} \label{sec:12}
\begin{thm} \label{thm:12}
Take system~\eqref{eq:dyn} with  inputs $u_{12}=1$, $u_{23}=0$ and
consider  the estimation of $\rho$ and $\Omega_{12}$ via~\eqref{eq:obs12}.
Take $\Gamma_{12}, \gamma_{12} >0$ and assume that  $\rho(0)$ and $\hat\rho(0)$ are real projectors  with $\tr{(P_1+P_2)\rho(0)}\in]0,1[$. Then,  for $\epsilon >0$ small enough, exists $\sigma >0$ such that, if  $1-\tr{\hat{\rho}(0)\rho(0)}\leq \sigma $ and  $|\hat{\Omega}_{12}(0)-\Omega_{12}|\leq \sigma$, then
 $\lim_{t\mapsto +\infty} \hat\rho(t)-\rho(t) = 0$ and $\lim_{t\mapsto +\infty} \hat{\Omega}_{12}(t)=\Omega_{12}$. Moreover the convergence is exponential.
\end{thm}

\begin{proof}  Since $\rho$ and $\hat\rho$ remain real projectors for $t>0$, they  can be seen as points on the two dimensional manifold $\mathbb{RP}^2$, a projective space. In particular~\eqref{eq:dyn} is a dynamical system with only $2$ degrees of freedom whereas the space of  $3\times 3$ symmetric real matrices where the dynamics is expressed  is of dimension $9$. We have try to use  less scalar variables but  averaging computations performed here below   are then much more complicated. In fact calculations based on  $3\times 3$ symmetric real matrices and thus with more variables than necessary  simplify notably  the analysis.

Set $\dotex \theta =\Omega_{12}$ and
consider the unitary and real transformation ($P_3=\ket{3}\bra{3}$)
$$
U(t)=\exp(\theta\sigma^{12} )= P_3 + \cos\theta(P_1+P_2) + \sin\theta \sigma^{12}
$$ and the attached change of frame  $\xi=U^\dag\rho U$, $\hat\xi=U^\dag\hat\rho U$.  Since
$$
U^\dag P_1 U = \frac{P_1+P_2}{2}+\frac{\cos(2\theta)}{2}\sigma_z^{12}+\frac{\sin(2\theta)}{2}\sigma_x^{12}
$$
and
$$
U^\dag \sigma_z^{12} U =
\cos(2\theta)\sigma_z^{12}+\sin(2\theta)\sigma_x^{12}
$$
system~\eqref{eq:obs12} reads
{\small
\begin{align*}
&\dotex{\hat\xi}=\epsilon\tilde\Omega_{12}[\sigma^{12},\hat\xi]+ ...
\\& ...\epsilon\Gamma_{12}\tr{
\left(\frac{P_1+P_2}{2}+\frac{\cos(2\theta)}{2}\sigma_z^{12}+\frac{\sin(2\theta)}{2}\sigma_x^{12}\right)
(\xi-\hat\xi)
}...\\& \quad
...\left(
\left(\cos(2\theta)\sigma_z^{12}+\sin(2\theta)\sigma_x^{12}\right) \hat\xi \right.
\\ & \qquad   +\hat\xi\left(\cos(2\theta)\sigma_z^{12}+\sin(2\theta)\sigma_x^{12}\right)
\\ & \qquad
\left.
-2\tr{\left(\cos(2\theta)\sigma_z^{12}+\sin(2\theta)\sigma_x^{12}\right)
\hat\xi}\hat\xi  \right)
\\
&\dotex\tilde{\Omega}_{12}=\epsilon\gamma_{12}
     \tr{\left(\cos(2\theta)\sigma_z^{12}+\sin(2\theta)\sigma_x^{12}\right)
     [\sigma^{12},\hat\xi]}...
     \\&\quad  ...
     \tr{\left(\frac{P_1+P_2}{2}+\frac{\cos(2\theta)}{2}\sigma_z^{12}+\frac{\sin(2\theta)}{2}\sigma_x^{12}\right)
       (\xi- \hat\xi)}
\end{align*} }
with  $\epsilon\tilde \Omega_{12}= \hat\Omega_{12}-\Omega_{12}$. Since $\dotex \xi =0$, for $\epsilon$ small enough  we can  consider the average system:
{\small
\begin{align*}
&\dotex{\hat\xi}=\epsilon\tilde\Omega_{12}[\sigma^{12},\hat\xi] + ...
 \\&\quad...
  \epsilon\frac{\Gamma_{12}}{4}\tr{\sigma_z^{12}(\xi-\hat\xi)}
\left(\sigma_z^{12} \hat\xi +\hat\xi\sigma_z^{12}
 -2\tr{\sigma_z^{12}\hat\xi}\hat\xi  \right) + ...
 \\& \quad
  ...\epsilon\frac{\Gamma_{12}}{4}\tr{\sigma_x^{12}(\xi-\hat\xi)}
\left(\sigma_x^{12} \hat\xi +\hat\xi\sigma_x^{12}
 -2\tr{\sigma_x^{12}\hat\xi}\hat\xi  \right)
\\
&\dotex\tilde{\Omega}_{12}=
  \epsilon\frac{\gamma_{12}}{4}\tr{\sigma_z^{12}[\sigma^{12},\hat\xi]}\tr{\sigma_z^{12}(\xi-\hat\xi)}+...
  \\& \quad
...\epsilon\frac{\gamma_{12}}{4}\tr{\sigma_x^{12}[\sigma^{12},\hat\xi]}\tr{\sigma_x^{12}(\xi-\hat\xi)}
\end{align*} }
But $\hat\xi=\xi$ and  $\tilde \Omega_{12}=0$ is a steady state of this average system. Assume we have proved that   this equilibrium is exponentially stable. Then the averaging theorem (see, e.g. \cite[theorem 4.1.1, page 168]{guckenheimer-holmes-book}) ensures that  the above time-periodic system admits a unique periodic orbit exponentially stable near $(\xi,0)$. Since $(\xi,0)$ is also an equilibrium of this time-periodic system, this exponentially stable orbit coincides with this equilibrium and the theorem is proved.

Let us prove now  that $(\xi,0)$ is a hyperbolically  stable equilibrium of the average system.
We have $\tr{\sigma_z^{12}[\sigma^{12},\hat\xi]}= \tr{[\sigma_z^{12},\sigma^{12}]\hat\xi}$ and
$[\sigma_z^{12},\sigma^{12}]=-2\sigma_x^{12}$.  Thus $\tr{\sigma_z^{12}[\sigma^{12},\hat\xi]}=2\tr{\sigma_x^{12}\hat \xi}$.
Similarly  $\tr{\sigma_x^{12}[\sigma^{12},\hat\xi]}=-2\tr{\sigma_z^{12}\hat \xi}$. Thus the average system reads
{\small
\begin{align}
&\dotex{\hat\xi}=\epsilon\tilde\Omega_{12}[\sigma^{12},\hat\xi] + ...\notag
 \\& \quad
  ...\epsilon\frac{\Gamma_{12}}{4}\tr{\sigma_z^{12}(\xi-\hat\xi)}
\left(\sigma_z^{12} \hat\xi +\hat\xi\sigma_z^{12}
 -2\tr{\sigma_z^{12}\hat\xi}\hat\xi  \right)  + ...\notag
 \\& \quad
  ...\epsilon\frac{\Gamma_{12}}{4}\tr{\sigma_x^{12}(\xi-\hat\xi)}
\left(\sigma_x^{12} \hat\xi +\hat\xi\sigma_x^{12}
 -2\tr{\sigma_x^{12}\hat\xi}\hat\xi  \right) \notag
\\
&\dotex\tilde{\Omega}_{12}=
  \epsilon\frac{\gamma_{12}}{2}
  \left(
    -\tr{\sigma_z^{12}\hat\xi}\tr{\sigma_x^{12}(\xi-\hat\xi)} + ... \right.\notag
    \\&\qquad \qquad\qquad\qquad ...\left.
    \tr{\sigma_x^{12}\hat\xi}\tr{\sigma_z^{12}(\xi-\hat\xi)}
    \right) \label{eq:average12}
\end{align} }
Assumption $\tr{(P_1+P_2)\rho(0)}>0$ implies that $\tr{(P_1+P_2)\xi}>0$.
We can choose the initial value of  $\theta$ such that $\tr{P_1\xi}=\tr{(P_1+P_2)\xi}>0$ and $\tr{P_2\xi}=0$.  $\xi$ and $\hat \xi$  belong to  $\mathbb{R P}^2$ and around  $\xi$ the variables
$
\hat x=\tr{\sigma_x^{12}\hat\xi}$ and $\hat z=\tr{P_1\hat\xi}$ form local coordinates for  $\hat \xi$: when $\hat \xi=\xi$, $\hat x=0$ and $\hat z = a$ with $a\in]0,1[$.  Some standard computations yield to the following linearized dynamics:
\begin{align}
    \dotex \tilde x&=-2\epsilon a \tilde\Omega_{12}-\frac{\epsilon a \Gamma_{12}}{2}\tilde x
    \notag
    \\
    \dotex \tilde z&=-\frac{\epsilon a(1-a)\Gamma_{12}}{2}\tilde z
    \notag
    \\
    \dotex\tilde{\Omega}_{12}&=\frac{\epsilon a\gamma_{12}}{2}\tilde x \label{eq:linav12}
\end{align}
with $\tilde x = \hat x $ and $\tilde z = \hat z-a$.  This linearized system is exponentially stable.

\end{proof}

\begin{rmk}
The stability of the  above average system~\eqref{eq:average12}  is more than  local.   It admits the following Lyapunov function:
$$ \frac{4}{\gamma_{12}}(\tilde\Omega_{12})^2
+\tr{\sigma_x^{12}(\hat \xi - \xi)}^2 + \tr{\sigma_z^{12}(\hat \xi - \xi)}^2
$$
Even if  theorem~\ref{thm:12} is a local stability result, the proposed estimator~\eqref{eq:obs12} should have a large attraction region. This is corroborated   by  simulations of figure~\ref{fig:perfect}.
\end{rmk}

\section{Estimation of $\Omega_{23}$} \label{sec:23}
\begin{thm} \label{thm:23}
Take system~\eqref{eq:dyn} with  inputs $u_{12}=1$, $u_{23}=\eta\cos\theta$ where  $\eta$  is constant  and  $\dotex \theta = \Omega_{12}$. Consider the estimation of $\rho$ and $\Omega_{23}$ via~\eqref{eq:obs23}.
Take $\Gamma_{23}>0$ and $\gamma_{23}>0$.
Assume $\rho$ is a real  projector with $\tr{(P_1+P_2)\rho(0)} >0$.
Then for $\epsilon$,  $\eta$ positive and  small enough, exists $\sigma >0$ such that, if $\hat\rho(0)$ is a real projector such that  $1-\tr{\hat{\rho}(0)\rho(0)}\leq \sigma $ and $|\hat{\Omega}_{23}(0)-\Omega_{23}|\leq \sigma$, then
 $\lim_{t\mapsto +\infty} \hat\rho(t)-\rho(t) = 0$ and $\lim_{t\mapsto +\infty} \hat{\Omega}_{23}(t)=\Omega_{23}$.
\end{thm}

\begin{proof}
The  unitary and real transformation $U=e^{\theta\sigma^{12}}$ reads
$ P_3 + \cos\theta~(P_1+P_2) + \sin\theta ~\sigma^{12}
$.
Consider the attached change of frame  $\xi=U^\dag\rho U$, $\hat\xi=U^\dag\hat\rho U$.
Since $ U^\dag\sigma^{23}U=\cos\theta~\sigma^{23}-\sin\theta~\sigma^{13}$, $\xi$ obeys to
$$
\dotex{\xi}=\eta\cos^2\theta ~\Omega_{23}[\sigma^{23},\xi]-
\eta\cos\theta\sin\theta~\Omega_{23}[\sigma^{13},\xi]
$$ and \eqref{eq:obs23} becomes
{\small
\begin{align*}
&\dotex{\hat\xi}=
\eta\cos^2\theta~\hat\Omega_{23}[\sigma^{23},\hat\xi]-\eta\cos\theta\sin\theta~\hat\Omega_{23}
[\sigma^{13},\hat\xi]+ ...\\
&\quad ...\epsilon\eta\Gamma_{23}\frac{1}{2}\tr{
\left(I_{12}+\cos(2\theta)\sigma_z^{12}+\sin(2\theta)\sigma_x^{12}\right)
(\xi-\hat\xi)
} ...
\\& \qquad\qquad\qquad ...(1-2\cos(2\theta))(\sigma_z^{23}\hat\xi+\hat\xi \sigma_z^{23}-2\text{Tr}(\sigma_z^{23}\hat\xi)\hat\xi)
\\
&\dotex{\hat{\Omega}}_{23}=\epsilon^2\eta\gamma_{23}\frac{1}{2}\tr{
\left(I_{12}+\cos(2\theta)\sigma_z^{12}+\sin(2\theta)\sigma_x^{12}\right)
(\xi-\hat\xi)
}
 ...
\\& \qquad\qquad\qquad ...(1-2\cos(2\theta))\text{Tr}(\sigma_z^{23}[\sigma^{23},\hat\xi])
\end{align*}}
In average,
$$
\tr{
\left(I_{12}+\cos(2\theta)\sigma_z^{12}+\sin(2\theta)\sigma_x^{12}\right)
(\xi-\hat\xi)
}(1-2\cos(2\theta))$$ is equal to $\tr{P_2(\xi-\hat\xi)}$.
After neglecting the highly oscillating terms, we obtain:
{\small
\begin{align*}
&\dotex{\xi}=\eta\frac{1}{2}\Omega_{23}[\sigma^{23},\xi]\\
&\dotex{\hat\xi}=\eta\frac{1}{2}\hat\Omega_{23}[\sigma^{23},\hat\xi]+ ...\\
&\quad ...\epsilon\eta\Gamma_{23}\frac{1}{2}\tr{P_2(\xi-\hat\xi)}(\sigma_z^{23}\hat\xi+\hat\xi \sigma_z^{23}-2\tr{\sigma_z^{23}\hat\xi}\hat\xi)\\
&\dotex{\hat{\Omega}}_{23}=\epsilon^2\eta\gamma_{23}\frac{1}{2}\tr{P_2(\xi-\hat\xi)}
\tr{\sigma_z^{23}[\sigma^{23},\hat\xi]}\\
\end{align*}}
In the time scale $\eta t$ instead of $t$ and up to a circular permutation   $(2,3,1)$ to  $(1,2,3)$, we recover~\eqref{eq:obs12} of theorem~\ref{thm:12}. We can always choose the initial value of  $\theta$ such $\tr{P_1\xi}=\tr{P_2\xi} = \frac{\tr{(P_1+P_2)\xi} }{2} >0$. Thus assumptions of theorem~\ref{thm:12} are satisfied, in particular $\tr{(P_2+P_3)\xi(0)} \in]0,1[$. Consequently,  for $\epsilon$ small enough, $(\hat \xi,\hat\Omega_{23})$ solution of the above average system converges locally exponentially towards $(\xi,\Omega_{23})$. Since $(\hat \xi,\hat\Omega_{23})=(\xi,\Omega_{23})$ is also  solution of the original system~\eqref{eq:obs23}, this implies that, for $\eta$ small enough,  $(\hat \xi,\hat\Omega_{23})$ converges  locally towards~$(\xi,\Omega_{23})$. This convergence is exponential.
\end{proof}

\section{Conclusion}

For a $3$-level system~\eqref{eq:dyn} with only a single population measurement we have proposed an algorithm in two steps for the estimation of $\Omega_{12}$ and $\Omega_{23}$. Simulations show the robustness  to additive noise of this algorithm relying on nonlinear asymptotic observers preserving the usual symmetries (change of frames). Theorems~\ref{thm:12} and~\ref{thm:23} ensure the local and exponential convergence. We can imagine switching periodically  between estimation of $\Omega_{12}$ via~\eqref{eq:obs12}  and estimation of $\Omega_{23}$ via~\eqref{eq:obs23} in order to produce  estimations of $\Omega_{12}$ and $\Omega_{23}$ in real-time.


\end{document}